\documentclass[usenatbib]{mnras}

\usepackage[T1]{fontenc}
\usepackage[utf8]{inputenc}
\usepackage{ae,aecompl}
\usepackage[usenames, dvipsnames]{color}

\usepackage{graphicx}	\usepackage{amsmath}	

\newcommand{\pc}{\,\mathrm{ pc }}

\newcommand{\Msun}{\,\mathrm{ M}_{\odot} }
\newcommand{\Rsun}{\,\mathrm{ R}_{\odot} }
\newcommand{\kpc}{\,\mathrm{ kpc }}
\newcommand{\Myr}{\,\mathrm{ Myr }}
\newcommand{\Gyr}{\,\mathrm{ Gyr }}
\newcommand{\yr}{\,\mathrm{ yr }}
\newcommand{\kms}{\,\mathrm{km\ s}^{-1}}

\newcommand{\erg}{\,\mathrm{ erg }}
\newcommand{\K}{\,\mathrm{ K }}

\renewcommand{\d}{\, \mathrm{d}}

\usepackage{commath}

\newcommand{\rc}{\mathrm{c}}

\newcommand{\rd}{\mathrm{d}}

\newcommand{\Ms}{m_{\star}}
\newcommand{\Rs}{R_{\star}}
\newcommand{\Mbh}{M_{\bullet}}
\newcommand{\rt}{r_{\mathrm{t}}}
\newcommand{\Jlc}{L_{\mathrm{lc}}}
\newcommand{\Jc}{L_{\rc}}

\usepackage{acronym}

\newacro{MBH}{massive black hole}
\newacroplural{MBH}[MBHs]{massive black holes}
\newacro{BH}{massive black hole}
\newacroplural{BH}[BHs]{massive black holes}

\newcommand{\BH}{\ac{BH}}
\newcommand{\BHs}{\acp{BH}}
\newcommand{\BHI}{BH1}
\newcommand{\BHII}{BH2}

\newacro{TDE}{tidal disruption event}
\newacroplural{TDE}[TDEs]{tidal disruption event} 
\newcommand{\TDE}{\ac{TDE}}
\newcommand{\TDEs}{\acp{TDE}}

\newcommand{\hugo}[1]{#1}

\title[TDEs in galaxy merger remnants]{Tidal disruption event rates in galaxy merger remnants}

\author[H. Pfister et al.]{Hugo Pfister,$^{1}$\thanks{E-mail: pfister@iap.fr}
Ben Bar-Or,$^{2}$
Marta Volonteri,$^{1}$
Yohan Dubois$^{1}$
and
Pedro~R. Capelo$^{3}$
\\
$^{1}$Sorbonne Universit\'{e}s, UPMC Universit\'{e} Paris 06 et CNRS, UMR7095,\\ Institut d'Astrophysique de Paris, 98bis boulevard Arago, F-75014, Paris, France\\
$^{2}$Institute for Advanced Studies, Einstein Drive, Princeton, NJ 08540, USA\\
$^{3}$Center for Theoretical Astrophysics and Cosmology, Institute for Computational Science, University of Zurich,\\
Winterthurerstrasse 190, CH-8057 Z$\ddot{u}$rich, Switzerland
}

\date{Accepted XXX. Received YYY; in original form ZZZ}

\pubyear{2019}

\begin{document}
\label{firstpage}
\pagerange{\pageref{firstpage}--\pageref{lastpage}}
\maketitle

\acused{TDE} \acused{BH} \acused{MBH} \begin{abstract}
The rate of tidal disruption events (TDEs) depends sensitively on the
stellar properties of the central galactic regions. Simulations show that
galaxy mergers cause gas inflows, triggering nuclear starbursts, increasing the central stellar density. Motivated by these numerical results,
and by the observed over-representation of post-starburst galaxies among
TDE hosts, we study the evolution of the TDE rate in high-resolution hydrodynamical
simulations of a galaxy merger, in which we capture the evolution of the stellar
density around the massive black holes (BHs). We apply analytical estimates of the loss-cone theory, using the stellar density profiles from simulations, to estimate the time evolution of the TDE rate. At the second pericentre, a nuclear starburst
enhances the stellar density around the BH in the least massive
galaxy, leading to an enhancement of the TDE rate around the secondary
BH, although the magnitude and the duration of the increase depend on the
stochasticity of star formation on very small scales.  The central stellar
density around the primary BH remains instead fairly
constant, and so is its TDE rate. After the formation of the
binary, the stellar density decreases, and so does the TDE rate.
\end{abstract}


\section{Introduction}

\acresetall{}
\acused{TDE} 

When a star passes sufficiently close to a \BH, it can get accreted. For
solar-type stars and \BHs\ with mass up to~$\sim$$10^8 \Msun$, the star is not
swallowed whole, but it is tidally perturbed and destroyed, with a fraction of its mass falling back on to the \BH\ causing a bright flare, known as
a tidal disruption event \citep[TDE;][]{Hills1975, Rees1988}.

A growing body of evidence suggests that \TDEs\ are
more likely to occur in host galaxies associated with recent starbursts
\citep{Arcavi+2014, French+2016, Stone_16a, Stone_16b,
  French+2017,Law-Smith+2017, Graur+2018}: the \TDE\ rate in these galaxies can be 30--200 times higher than in main-sequence
galaxies, with galaxy mergers a possible cause for the starburst~\citep{Zabludoff+1996, Yang+2004, Yang+2008,2009MNRAS.395..144W}. \cite{Stone_16b} advanced the hypothesis that this increase could be due to an anomalously high central stellar density, from which most \TDEs\ are sourced, caused by the starburst. To test this hypothesis, we set ourselves in a case including a strong nuclear starburst: a galaxy merger, when gas inflows due to tidal forces and ram-pressure shocks can trigger nuclear starbursts that form a dense stellar cusp and temporarily increase the central density~\citep{Mihos_96,  VanWassenhove_14,Capelo_17,2018MNRAS.tmp.1993S}.~\cite{VanWassenhove_14} find an enhancement of almost two orders of magnitude of the density within 10 pc around the secondary BH of a 1:4 merger, during the 150~Myr following the starburst. This suggests that, during the merger, the \TDE\ rate can increase by a few orders of magnitude.

\section{A lower limit for the tidal disruption event rate}
\label{tde_rate}
In this section, we perform an approximate  calculation to understand what are the physical parameters affecting the TDE rate $\Gamma$, defined as the number of disruptions per galaxy per unit time. In practice, for the rest of this work, we estimate the TDE rate with a more elaborated method detailed in \S\ref{sec:TDERateInOurSimulation}.

Stars of mass $\Ms$ and radius $\Rs$ are disrupted if the pericentre distance
to the \BH, of mass $\Mbh$, is smaller than the tidal disruption radius
$\rt\sim{(\Mbh/\Ms)}^{1/3}\Rs$.  This defines a ``loss
cone''~\citep{Lightman+1977} in angular momentum of size $\Jlc^2/\Jc^2(E)$,
where $\Jlc=\sqrt{2 G \Mbh\rt}$ is the maximal angular momentum per unit mass for disruption,
$G$ the gravitational constant, $\Jc(E)$ is the circular (maximal) angular
momentum per unit mass of an orbit, with energy per unit mass $E= v^2/2 + \Phi(r)$, $\Phi(r)$
is the gravitational potential, and $r$ and $v$ are, respectively, the distance
to the \BH\ and relative speed.

It is customary to define two regions, whose contributions to the flux of stars
match at the critical radius $r_\rc$, with the corresponding specific energy
$E_\rc=\Phi({r_\rc})$~\citep{Syer+1999}. The first is a region close to the
\BH\ ($E<E_\rc$, $r<r_\rc$), where the time to diffuse across the loss cone is
longer than the orbital period. All stars inside the loss cone will be
disrupted at periapsis and the loss cone is empty. Farther away from the
\BH\ ($E>E_\rc$, $r>r_\rc$), the time to diffuse across the loss cone is
shorter than the orbital period.  Stars will scatter in and out of the loss
cone during the orbital motion and the loss cone is full.

In the ``empty loss-cone'' region, the \TDE\ flux (events per unit time per
unit energy), depends only logarithmically
on the size of the loss cone and it is given by~\citep[e.g.][]{Magorrian+1999,
  Wang+04}:
\begin{equation}
  \mathcal{F}_\mathrm{empty}(E) =  \frac{\mu N(E)}{\ln(\Jc^2/\Jlc^2)} 
  \sim  \frac{N(E)}{T_r \ln(\Jc^2/\Jlc^2)} \, ,
\end{equation}
where $N$ is the energy density function, $N(E)=4\pi^2 \Jc^2 f(E) P(E)$ for an ergodic phase-space distribution function $f(r,v)~=~f(E)$ \citep[e.g][]{Merrit_book}; $\mu$ is the orbit-averaged diffusion
  coefficient in angular momentum~\citep[see Eq.~(13c) in ][]{Vasiliev2017} and $T_r$ is the relaxation timescale \citep{Spitzer_58}.

Farther away from the \BH,\ in the ``full loss-cone'' region, the \TDE\ flux 
depends linearly on the size of the loss cone and it is given by:
\begin{equation}
  \label{eq:dGamma_flc}
   \mathcal{F}_\mathrm{full}(E) = \frac{N(E)}{P(E)} \frac{\Jlc^2}{\Jc^2},
\end{equation}
where $P$ is the radial period.

The total \TDE\ rate $\Gamma$ is the integral over these two fluxes:
\begin{equation}
  \label{eq:TDE_rate}
  \Gamma =  \int \!\! \mathcal{F}(E) \rd E  \sim \int_{-\infty}^{E_\rc}   \mathcal{F}_\mathrm{empty}(E)  \rd E +
  \int_{E_\rc}^{\infty}   \mathcal{F}_\mathrm{full}(E)  \rd E \, ,
\end{equation}
\hugo{where $E_c$ is defined such that $\mathcal{F}_\mathrm{empty}(E_\rc) =
\mathcal{F}_\mathrm{full}(E_\rc)$}. To carefully estimate the TDE rate, one should compute the two integrals. However, in practice, the density profile close to the \BH\ is unknown and in this work we consider only the region outside the critical radius. Therefore, we estimate a lower limit to the TDE rate, considering only the full loss-cone regime:

\begin{equation}
  \label{eq:TDE_flc}
  \Gamma_\mathrm{full} =  4\pi^2 \Jlc^2 \int_{E_\rc}^{\infty} \rd E  f(E) = \pi
  \Jlc^2 \Ms^{-1} \rho(r_\rc)\langle {v}^{-1} \rangle (r_\rc) \, ,
\end{equation}
where $\rho$ is the stellar density and $\langle v^{-1} \rangle (r)~=~m_\star \rho^{-1}(r) \int f(r,v) v^{-1}\d ^3 \vec{v}$. If we set $\langle {v}^{-1}
\rangle\sim\sqrt{2/\pi}/\sigma$, with $\sigma = \sqrt{\langle v^2 \rangle/3}$
being the velocity dispersion, we  have:
\begin{align}
  \label{eq:TDE_rho_sig}
  \Gamma_\textrm{full} \sim {} 
  &
    5.0 \times 10^{-4} \yr^{-1}\times \\
    & {\left(\frac{\Mbh / \Ms}{10^6}\right)}^{4/3}
    \frac{\Rs}{\Rsun} \frac{\rho(r_\rc)}{10^6 \Msun \pc^{-3}} \frac{100 \kms}{\sigma(r_\rc)}. \nonumber
\end{align}
We can obtain $r_\rc$ by equating the full and empty loss-cone fluxes, this yields:
\begin{equation}
  \label{eq:r_crit1}
\frac{G M(r_\rc)}{\sigma^2(r_\rc)} \sim \frac{4\sqrt{2}}{3 \pi} \,{\left(\frac{\Mbh}{\Ms}\right)}^{4/3} \Rs,
\end{equation}
where we have assumed that the enclosed stellar mass within $r$, $M(r)$,  equals
$4 \pi\rho(r) r^3/3 $. To get a step further, we assume that ${\sigma^2(r) \sim G(M_\bullet + M(r))/r}$, and that $M(r_\rc) \sim \Mbh$. \hugo{Note that this last assumption is not necessarily true, but happens to give excellent results in our case (see \S\ref{sec:TDERateInOurSimulation}).} This yields:
\begin{equation}
  \label{eq:r_crit}
	r_\rc \sim 3 \pc \,{\left(\frac{\Mbh/\Ms}{10^6}\right)}^{4/3} \,{\left(\frac{\Rs}{\Rsun}\right)} \,.
\end{equation}

During a galaxy merger, $\rho(r_\rc)$ can change by orders of
magnitude~\citep{VanWassenhove_14}, while there is only moderate change in
$M_\bullet$ and $\sigma$ (and, consequently, $r_\rc$). Therefore, our limit to the \TDE\
rate depends, almost exclusively, on the density at the radius $r_\rc$, which depends only on the BH mass, for stars with similar mass and radius. This calculation is presented to understand the physical parameters impacting the TDE rate. We describe the method we use to estimate $\Gamma$ in \S\ref{sec:TDERateInOurSimulation}.

\section{Simulations}
\label{section:Simulation}

Similarly to~\cite{Pfister_17}, we perform a zoom re-simulation of the 1:4
coplanar, prograde--prograde galaxy merger from~\cite{Capelo_15}, which was
shown to have a strong burst of nuclear star formation~\citep[see
also][]{VanWassenhove_14}, and is adopted here as a reference merger to highlight
the various physical processes responsible for the evolution of the
nucleus.  
Similar bursts were also observed in mergers with mass ratio 1:2
(coplanar and inclined orbital configurations), whereas lower mass-ratio
mergers had weaker (1:6 case) or negligible (1:10) nuclear starbursts.
Initially \BHI, with a mass of $3.53\times 10^6 \Msun$, is in the main galaxy,
whereas \BHII, with a mass of $0.88\times 10^6 \Msun$, is in the secondary
galaxy.

We re-simulate the merger phase \citep[see][]{Capelo_15}, which begins at the
second pericentre, at ${t\sim 1\Gyr}$, and lasts until the binary \BH\ has
formed, 300~Myr later. It is
during this phase that the starburst occurs and we expect variations in the
density and, consequently, in the \TDE\ rate.

This re-simulation (\texttt{Resim0}) is performed with the public code
\texttt{Ramses}~\citep{Teyssier_02}. \texttt{Ramses} is an adaptive mesh
refinement code in which the evolution of the gas is followed using a
second-order unsplit Godunov scheme for the Euler equation. The approximate
Harten--Lax--Van Leer Contact \citep{Toro_97} Riemann solver with a MinMod total
variation diminishing scheme to reconstruct the interpolated variables from
their cell-centred values is used to compute fluxes at cell
interfaces. Collisionless particles, dark matter (DM), stellar, and \BH\
particles, are evolved using a particle-mesh solver with a cloud-in-cell (CIC)
interpolation. The mass of DM particles
($m_\mathrm{DM} = 1.1\times10^5\Msun$) and stellar particles
($3.3\times 10^3\Msun$) is kept similar to that in~\cite{Capelo_15} but we
allow for better spatial resolution (down to $\Delta x=0.76 \pc$), refining the
mesh where
$M_\mathrm{DM}^\mathrm{cell}+ 10 M_{\rm b}^\mathrm{cell} \geq 8 m_\mathrm{DM}
$, where $M_\mathrm{DM}$ and $M_{\rm b}^\mathrm{cell} $ are, respectively, the
mass of DM and baryons in the cell. Maximum refinement is enforced within
$4\Delta x$ around the \BH\@.

\begin{figure*}
\includegraphics[width=0.5\columnwidth]{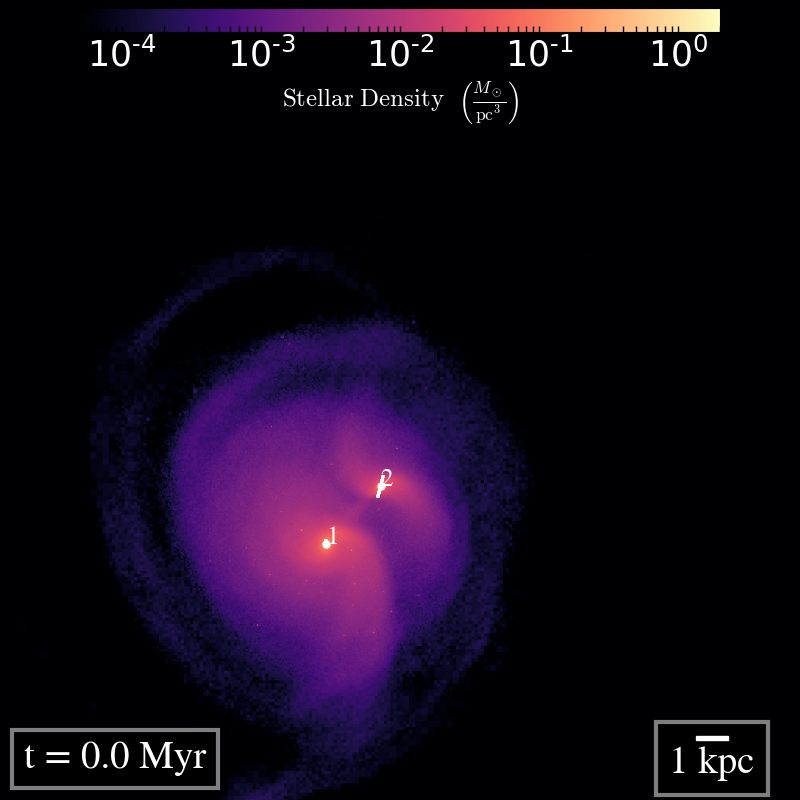} 
\includegraphics[width=0.5\columnwidth]{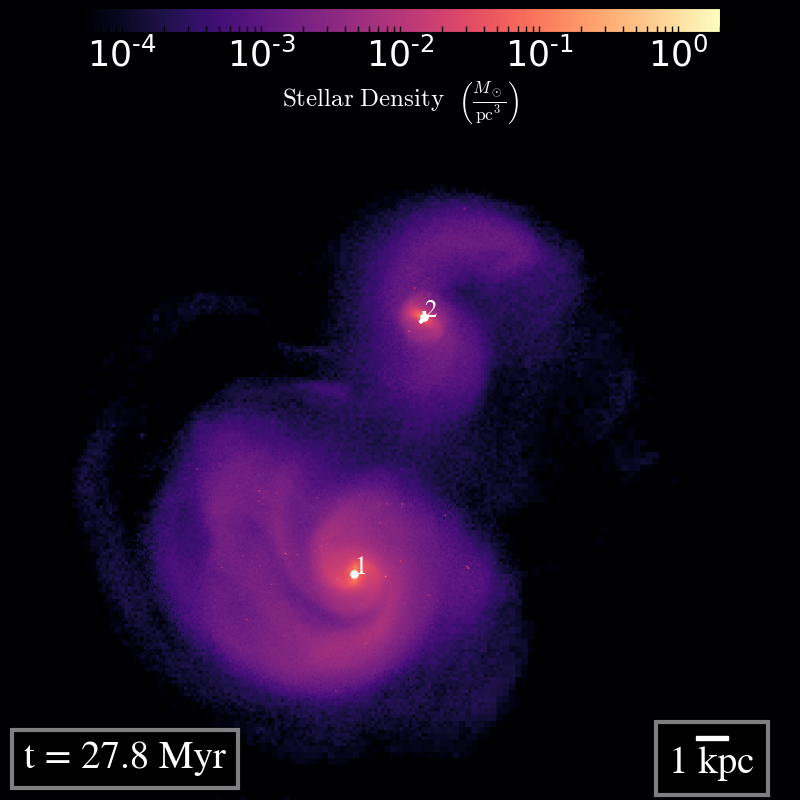} 
\includegraphics[width=0.5\columnwidth]{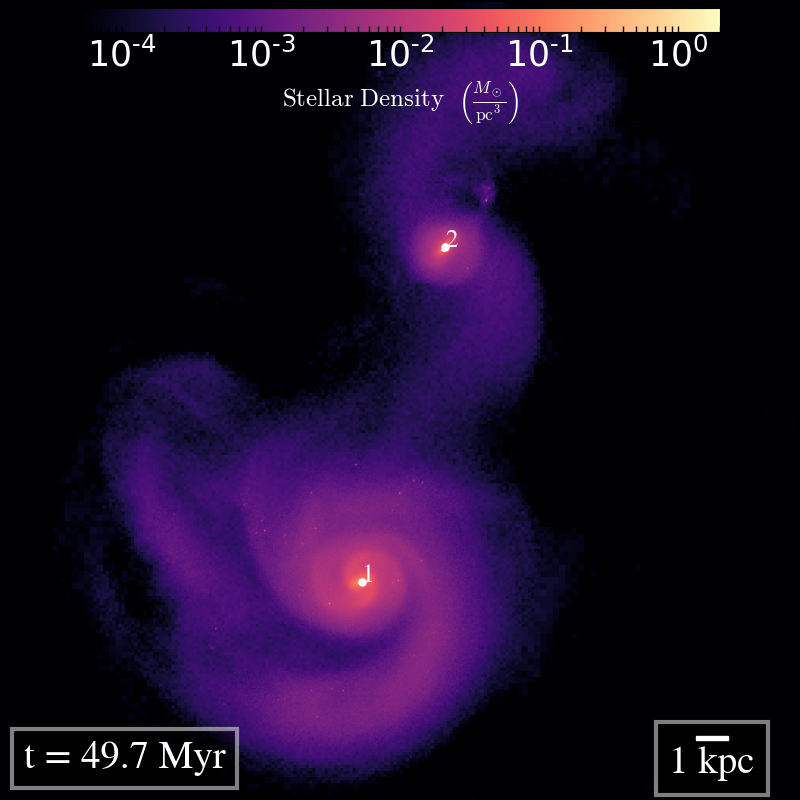}
\includegraphics[width=0.5\columnwidth]{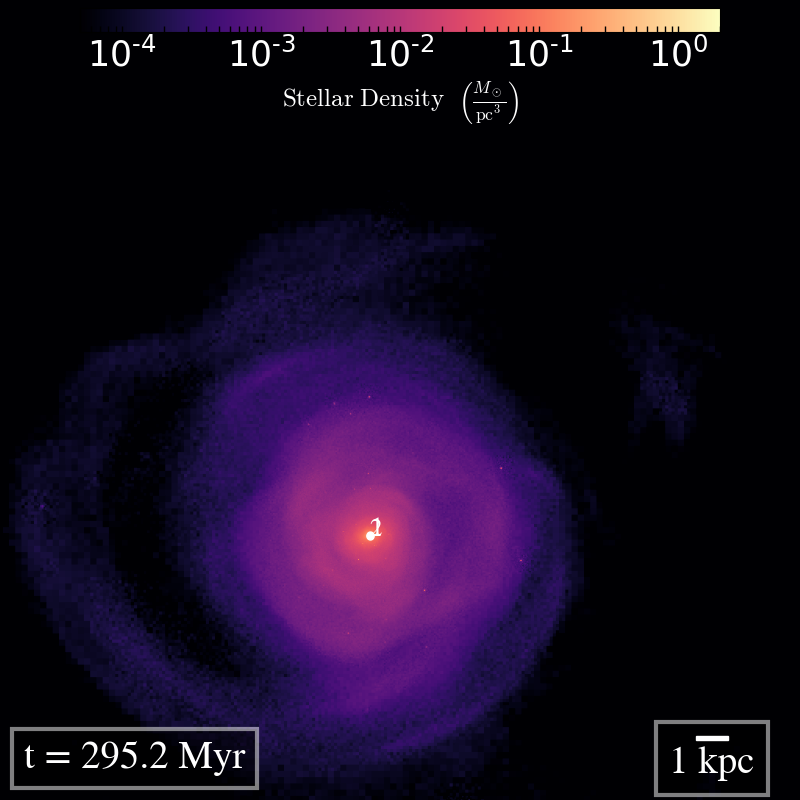} \\
\includegraphics[width=0.5\columnwidth]{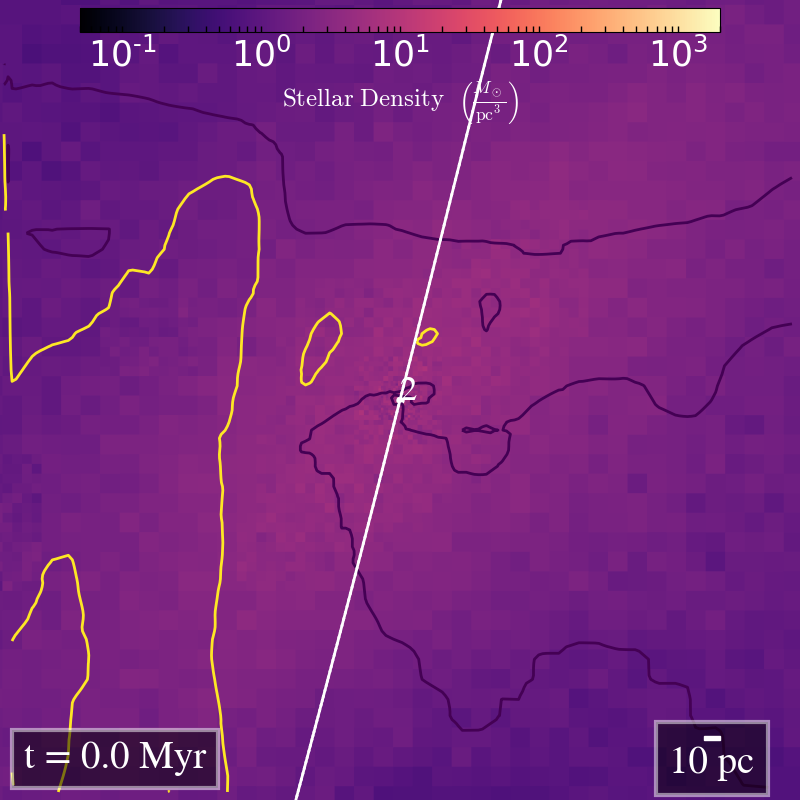} 
\includegraphics[width=0.5\columnwidth]{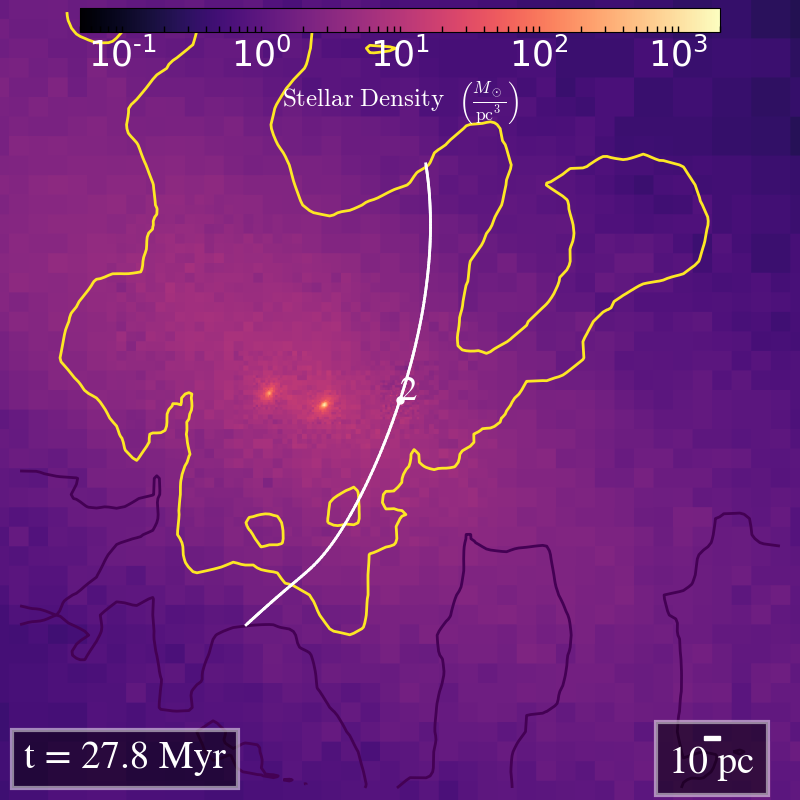}
\includegraphics[width=0.5\columnwidth]{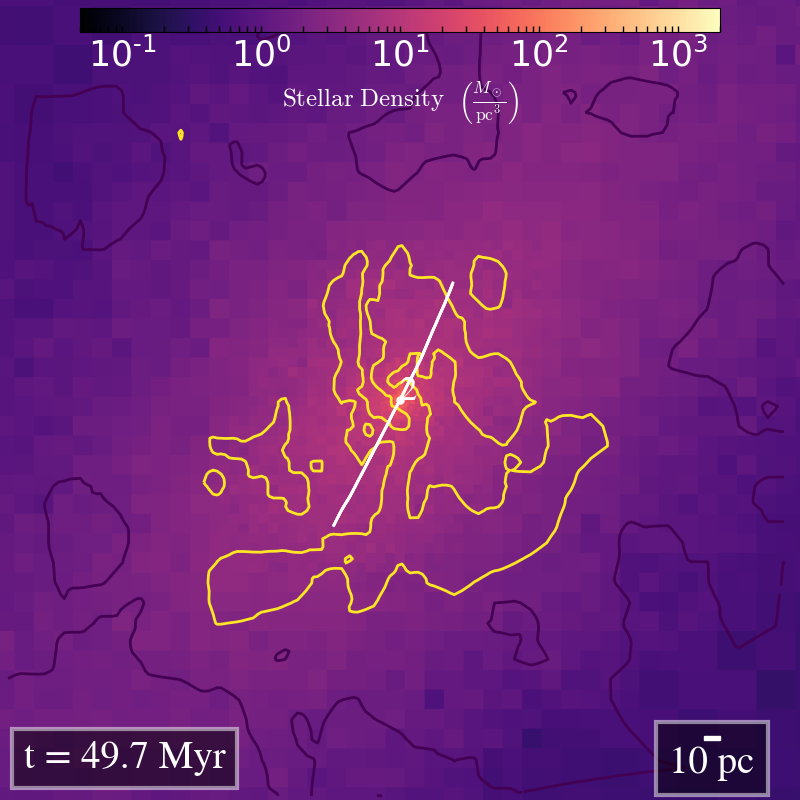} 
\includegraphics[width=0.5\columnwidth]{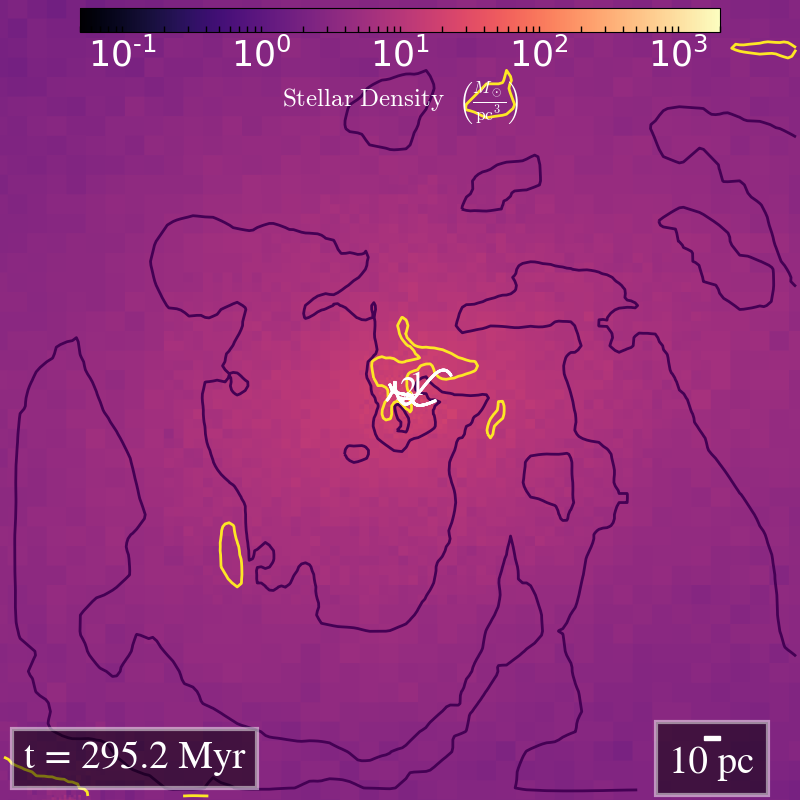}
\caption{Stellar density maps of the two galaxies (top row) and centred on the
  secondary \BH\ (bottom row). Initially, the \BH\ proceeds on a smooth
  trajectory (first column); then, the starburst occurs and some newly formed
  stellar clumps deviate the \BH\ from its smooth trajectory (second column);
  at some point, those clumps merge and the \BH\ gets trapped (third column);
  finally, the \BH\ binary forms in the remnant galaxy (fourth column). The
  white line in the bottom images represents the position of the \BHs\ within
  $\pm 1\Myr$. In order to show how irregular is the gas density compared with the stellar one, we indicate the iso-$\rho_{\rm gas}$ contours of 1 (10) a.m.u. cm$^{-3}$ with purple (yellow) lines.}
\label{fig:snapshots}
\end{figure*}

\begin{figure}
\centering
\includegraphics[width=\columnwidth]{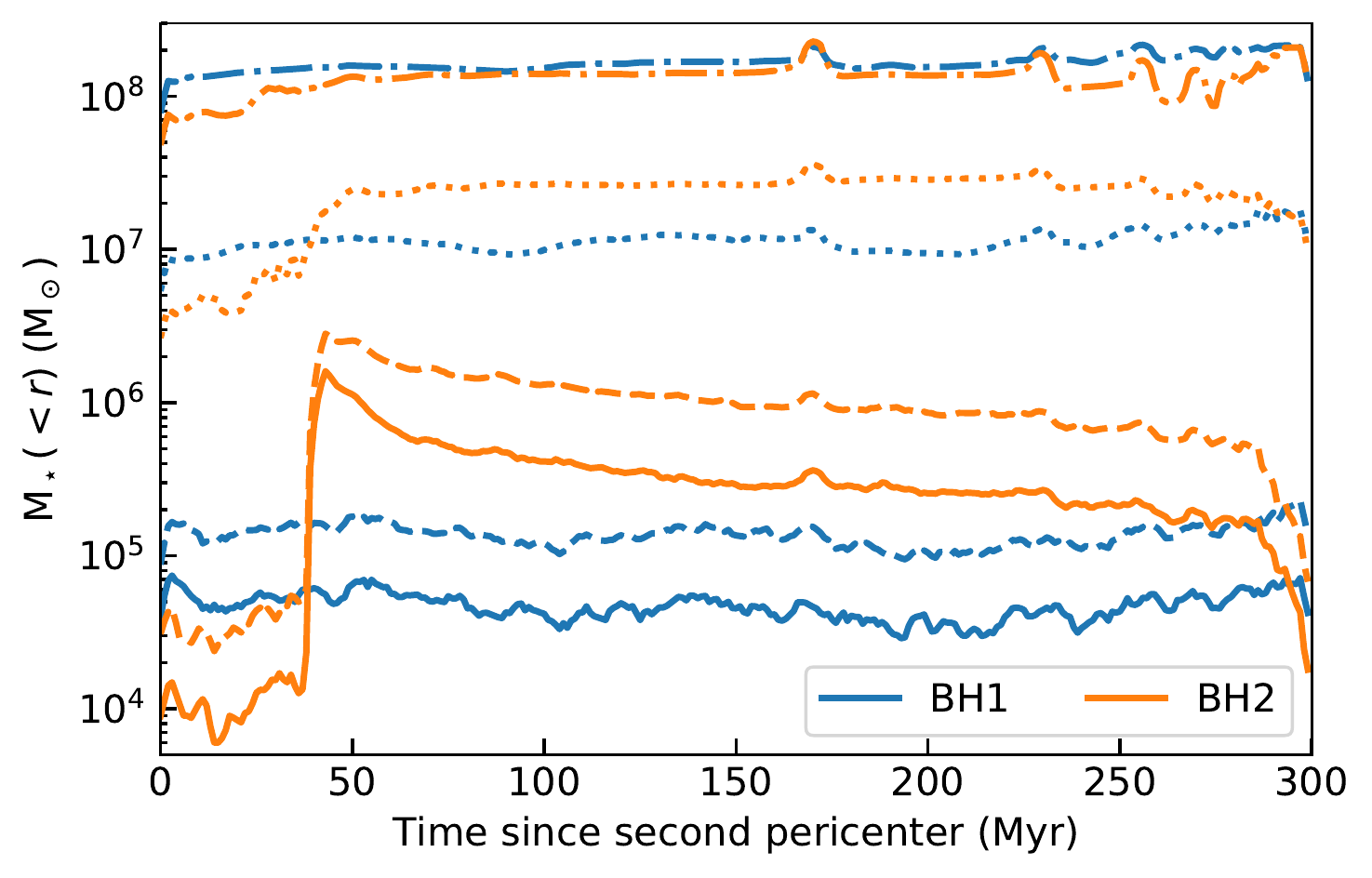}
\caption{Enclosed stellar mass within 3 (solid), 5 (dashed), 30 (dotted), and
  100 (dash-dotted) pc around each \BH, as a function of time elapsed since the
  second pericentre.}
\label{fig:MassEnc}
\end{figure}

\subsection{Physics of galaxies}
Gas is allowed to cool with the contribution of hydrogen, helium, and metals
using tabulated cooling rates from~\cite{Sutherland_93} above $10^4\K$, and
rates from~\cite{rosen&bregman95} below $10^4\K$ and down to $10 \K$.

Star formation, occurring at gas densities above $1\, \rm H\, cm^{-3}$, is
stochastically sampled from a random Poisson distribution~(see \citealt{Rasera_06} for details) following a Schmidt law for the local star formation rate
$\dot \rho = \epsilon\rho_{\rm gas}/t_{\rm ff}$, where $\rho$ and $\rho_{\rm gas}$ are
the stellar and gas density, respectively, $t_\mathrm{ff}$ is the local gas
free-fall time, and $\epsilon$ depends on the local gravo-turbulent properties
of the gas, as detailed in~\cite{Trebitsch_18}.

For the feedback from supernovae (SNe), we use the Sedov/snowplough-aware method
described in~\cite{Kimm_14}, in which stars release
$2\times10^{49}\erg\, \Msun^{-1}$ after $5\Myr$ (assuming 20 per cent of the mass of
star particles contributes to type II SNe).

\subsection{Physics of black holes}
We use the model of \BHs\ described in~\cite{Dubois_12}, where accretion is
computed using the Bondi--Hoyle--Lyttleton formalism capped at the Eddington
luminosity. BH feedback consists of a dual-mode approach, wherein thermal energy, corresponding to 15 per cent of the bolometric luminosity (with radiative efficiency $\epsilon_{\rm r} = 0.1$), is injected at high accretion rates (luminosity above 0.01 the Eddington luminosity); otherwise, feedback is modelled with a bipolar jet with a velocity of $10^4$~km~s$^{-1}$ and an efficiency of 100 per cent.

We modify the implementation of \BH\ dynamics. In~\cite{Dubois_12}, the
mass of the \BH\ is spread in a sphere of $4\Delta x$ radius around the \BH\ in
order to smooth the gravitational potential it generates. However, when two
\BHs\ approach each other, the formation of the binary is delayed. Here, we
deposit all the mass of each individual \BH\ on the particle before performing
the CIC interpolation, to obtain more accurate dynamics.

\subsection{\TDE\ rate in the simulation}
\label{sec:TDERateInOurSimulation}

In Section~\ref{tde_rate}, we derived Eq.~\eqref{eq:TDE_rho_sig} to get a
  physical insight of the relevant parameters affecting the \TDE\ rate. In
  practice, however, we measure the stellar density profiles around BHs for
  each snapshot in our simulation and fit them with a double power-law profile
  $\rho(r) = \rho_0r^{\gamma}{(1+r/r_0)}^{\beta-\gamma}$. We then pass these
  density profiles to the \texttt{PhaseFlow} code (included in \texttt{Agama};
  \citealt{Vasiliev2017,Vasiliev2018}) which Eddington inverses them to obtain
  the density function $f(E)$, and compute the loss-cone filling factor
  $q(E) = \mu P(E) \Jc^2/\Jlc^2 = \mathcal{F}_\mathrm{empty} /
  \mathcal{F}_\mathrm{full} \ln(\Jc/\Jlc)$. The \texttt{PhaseFlow} code is conceived to solve
  the time-dependent Fokker--Planck equation, but we only use it to estimate $f$ and $q$ at each timestep corresponding to a snapshot of the simulation.
  
\cite{Cohn+1978} estimated the instantaneous TDE flux per unit time and
  energy $\mathcal{F}$ (see Eq.~(10--13) in~\cite{Wang+04} or Eq.~(16--17) in \cite{Stone_16a}). We use a
  slightly modified version of this approximation (see Eq.~(14) in
  \cite{Vasiliev2017}):
  \begin{align}
  \mathcal{F}(E) = \frac{q(E) L^2_{lc}/L_c^2 }{(q(E)^2+q(E)^4)^{1/4}+\ln(L_c^2/L_{lc}^2)}  \frac{N(E)}{P(E)} \, .
  \end{align}
  This expression can be integrated to obtain the TDE rate
  $\Gamma$. From $q$, we can also estimate the critical radius/energy solving
  $q(E_c)=q(\Phi(r_\rc))=\ln(\Jc/\Jlc)$. Using this technique, we found that $r_c$ is about 20 pc at all times for BH1, and is initially 13 pc for BH2, but drops to 4--5 pc after the starburst (see \S\ref{sec:CuspEvolutionDuringTheMerger}). These numbers are in very good agreement with the estimates of Eq.~\eqref{eq:r_crit}: 22 and 5 pc for BH1 and BH2 (their masses are respectively $4.4\times 10^6\Msun$ and $1.4\times 10^6 \Msun$, almost constant during the simulation). Consequently, for the rest of the paper, we adopt our approximate estimate of $r_c$.
\section{Results}

\subsection{Nuclear starburst}
\label{sec:CuspEvolutionDuringTheMerger}

In Fig.~\ref{fig:snapshots}, we show stellar density maps of our simulation. In Fig.~\ref{fig:MassEnc},
we show the enclosed stellar mass around each \BH\ as a function of time, for different
radii in the re-simulation. It is clear from Fig.~\ref{fig:MassEnc} that the
primary galaxy is not affected by the merger: during the ${300 \Myr}$ of the
simulation, very few stars form around the primary BH, in agreement with the lower-resolution run~\citep{Capelo_15}. Therefore, the
\TDE\ rate should remain roughly constant.

\begin{figure}
\centering
\includegraphics[width=\columnwidth]{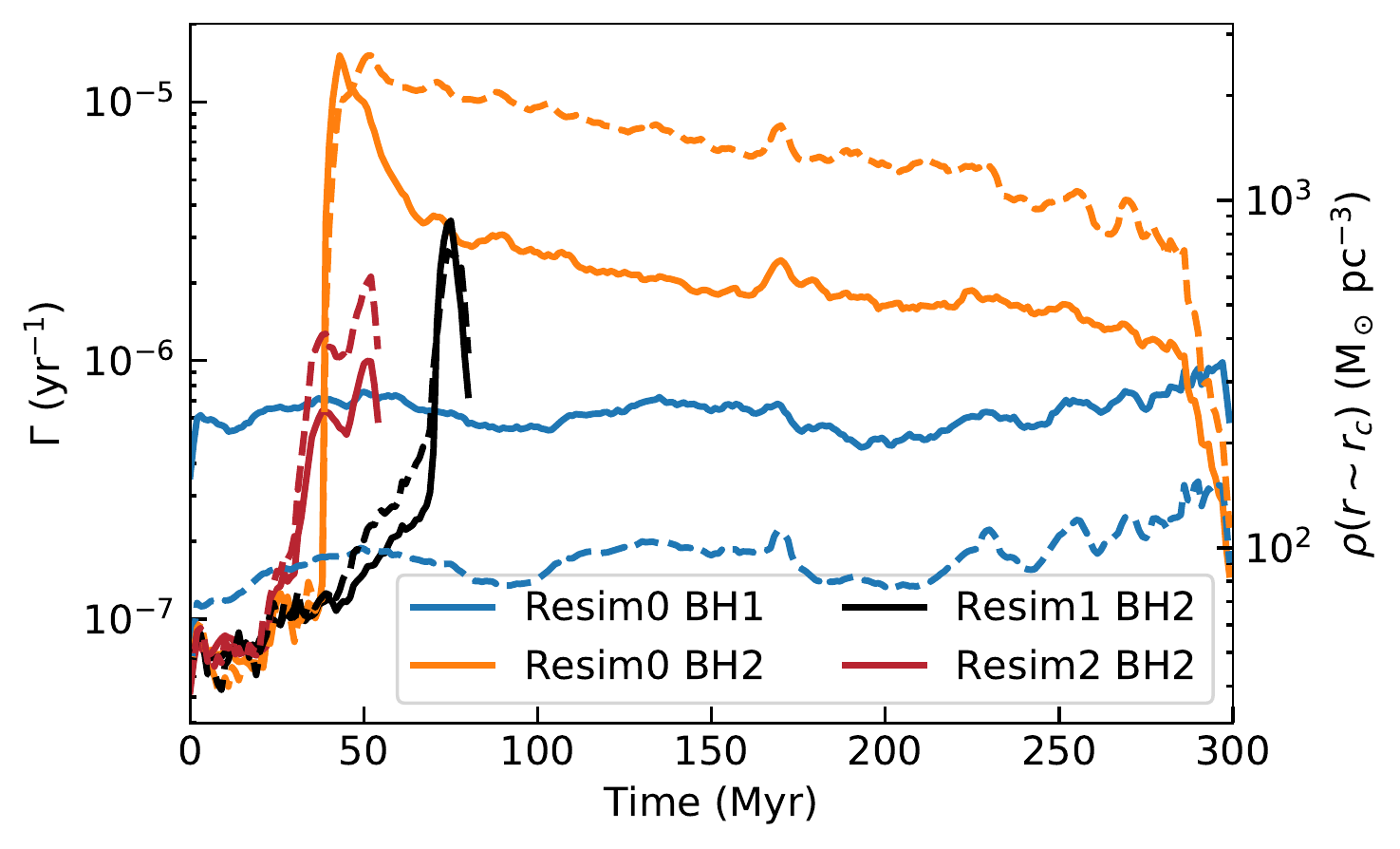}
\caption{TDE rate around each \BH\ (solid line) and stellar density at the
  critical radius (dashed line). $r_c$ is estimated from Eq.~\eqref{eq:r_crit}: as the masses of BH1 and BH2 are respectively $4.4\times 10^6$ and $1.4\times 10^6 \Msun$, their respective $r_c$ are $22\pc$ and $5 \pc$. We show the same quantities for BH2 in the other re-simulations
  (see Section~\ref{sec:EffectOfSto}), which are run for a shorter time as we
  are only interested in the enhancement of the stellar density following the
  first starburst. All quantities are shown as a function of time.}
\label{fig:TDE}
\end{figure}

The secondary galaxy, instead, undergoes a major starburst just after
the second pericentre, lasting~${50\Myr}$. As the gas is perturbed by tidal torques and ram-pressure shocks, it loses angular momentum and falls towards the centre, triggering nuclear star formation. In
the original simulation from~\cite{Capelo_15}, this first burst is followed by
other bursts similar in magnitude \citep[see the left-hand panel of Fig.~1 in][]{Capelo_15} that we do not see in the
re-simulation. The main reason is the increase of resolution, which results in higher gas density, causing initially elevated levels of nuclear star formation, with respect to the lower-resolution run, which consume a fraction of the accumulating gas. Another difference
with the original simulation from \cite{Capelo_15} is that we use a more
physically motivated model for star formation, with a variable star formation
efficiency: the star formation rate, therefore, is not directly proportional to
the gas density. Furthermore, our much higher resolution results in clumpy star
formation, as shown in the second column of Fig.~\ref{fig:snapshots}. These
clumps are fairly small (few pc size) but can be very massive, up to a few
$10^6 \Msun$, similar to the mass of \BHII~($\sim 1.4\times 10^6 \Msun$). This leads
to interactions that scatter the \BH\@. Consequently, the density ``seen'' by
the \BH\ is highly dependent on local stochastic processes. The enclosed mass
within ${5\pc}$ from \BHII\ (orange dashed line in Fig.~\ref{fig:MassEnc}) is
almost constant, until it increases abruptly as the clumps merge and capture
the \BH\ at about 50 Myr. This is clear both from the third column of
Fig.~\ref{fig:snapshots} and from Fig.~\ref{fig:MassEnc}. After this rise in
density, the enclosed mass within 5 pc does not vary until the binary forms,
whereas the enclosed mass within 3 pc decreases. This is contrary to the expectations of the evolution of a mass distribution around a BH, which normally contracts \citep{1976ApJ...209..214B,1995ApJ...440..554Q}. However, at difference with the assumptions in classic approaches, which look at equilibrium, steady-state solutions or BHs growing slowly within the stellar distribution,  the BH enters rapidly the stellar clump, and the mass of the clump and the BH are similar. The effect we observe can be explained assuming that the system BH-clumps suffers a series of high-speed encounters \citep{BT_87}, bringing enough energy to start the disruption of the clump, although we cannot rule out that the effect is numerical.
When the binary forms, \emph{i.e.} the BHs are separated by about 1 pc, the enclosed mass
decreases again. This might be due to heating: when the binary shrinks, it releases
energy in the nucleus. Since the simulation cannot resolve scatterings between
stars and the binary, we are unable to rigorously confirm if this effect is
physical or a numerical artifact, although detailed $N$-body simulations show
similar results~\citep[e.g.][]{2001ApJ...563...34M}.

In summary, the amount of stars around \BHII\ changes significantly during the
merger, and thus we expect large variations of its \TDE\ rate. However, the
exact enhancement may depend on the position of the \BH, which can be chaotic due to three-body interactions with stellar clumps. The amount of stars around \BHI\ remains fairly constant and we do not expect much change in the
\TDE\ rate until it binds with \BHII\ and it is embedded in the same stellar
environment.

\subsection{TDE rate}
\label{sec:TDERate}
Using the techniques described in Section~\ref{sec:TDERateInOurSimulation}, we
estimate the \TDE\ rate as a function of time in the simulations. Note that
here we have taken the conservative assumption of not including an inner cusp
around the \BHs\ \citep{1976ApJ...209..214B}, hence the estimated \TDE\ rate is a lower bound.

We show in Fig.~\ref{fig:TDE}, as a function of time, the \TDE\ rate around
each \BH\ (solid line) and the density at the critical radius (dashed line), as
defined in Eq.~\eqref{eq:r_crit}. Note the remarkable
agreement between the \TDE\ rate measured with the \texttt{PhaseFlow} code and the
stellar density at $r_\rc$. 

The initial \TDE\ rate is very small ($\sim$$10^{-7}\yr^{-1}$ for both \BHs),
because the density around each \BH\ is very low: we find, for the two \BHs, a
stellar density of $\sim$$10^2\Msun\, \pc^{-3}$, which is one to two orders of
magnitude lower than in local galaxies~\citep{1997AJ....114.1771F}. The reason is
that the analytical initial conditions of the merging galaxies \citep{Capelo_15}
assume that the stellar bulge is described by a spherical Hernquist profile \citep{Hernquist1990} with inner logarithmic slope $\gamma=-1$, whereas local galaxies exhibit a range of inner density slopes going from $\gamma \sim 0$ to $\gamma=-2$
\citep{1997AJ....114.1771F,2007ApJ...664..226L}, up to $\gamma=-4$ in the presence of nuclear star clusters, common in low-mass
galaxies \citep{2011ApJ...726...31G}. In addition, before the beginning of the merger
simulation, galaxies are relaxed for ${100\Myr}$ and, during this time, the velocity distribution
near the resolution limit (${10 \pc}$) is not well sampled because of the limited
number of stars, leading to an even shallower profile than the initial
Hernquist profile.

The \TDE\ rate around \BHI\ is fairly constant, irrespective of the dynamical
phase of the merger: since the stellar density profile around \BHI\ is not affected by
the merger, the amount of stars available to be disrupted is constant and so is
the \TDE\ rate. The \TDE\ rate around \BHII\ is instead increased by a factor of about 30 during the $250\Myr$ following the burst, with a short peak of more than two orders of magnitude enhancement. During the first 200~Myr of this enhancement, the two galaxies can be separated by more than 1~kpc, up to 10~kpc.  While the maximum value of $\sim$$10^{-5} \yr^{-1}$ may seem surprisingly low, we recall that the initial density profile, after relaxing the initial conditions, was shallow  and we do not include the possibility of a stellar cusp due to unresolved stellar dynamics, which would increase the initial TDE rate and, perhaps, decrease the relative enhancement caused by merger-driven nuclear star formation.

As discussed in Section~\ref{sec:CuspEvolutionDuringTheMerger}, the central
density and the \TDE\ rate drop once the binary is formed.  However, to calculate the TDE rate we assumed a single \BH\@ surrounded by a spherical density distribution, which is not valid any longer after formation  of the binary. More sophisticated techniques, beyond the scope of this paper, can be used for binary BHs \citep[e.g.][]{2019arXiv190104508L}, which often result in an increased rate, at least for a short time \citep[e.g.][]{Chen_09,2011ApJ...729...13C, Li+2017}.

\subsection{Effect of stochasticity}
\label{sec:EffectOfSto}

We rerun the exact same simulation, but changing the random seed used in the stochastic sampling of star formation (\texttt{Resim1} and \texttt{Resim2}), and perform the same analysis. This test is done for three main reasons: firstly, reproducibility of our results; secondly, the small number of particles around the BH in the early phase before the starburst (about $10^4 \Msun$ within 3 pc, corresponding to 10 stellar particles; see Fig.~\ref{fig:MassEnc}) might affects our results; thirdly, because
reaching pc-resolution is a double-edged sword. On the one hand, we
resolve the gas flows and star formation very close to the \BH\@. On the other
hand, the stochasticity of very local processes becomes important. The exact
position and mass of the forming stellar clumps have strong effects on both the
orbits of \BHs\ and on the density around them.

We show in Fig.~\ref{fig:TDE} the \TDE\ rate and density at the critical radius around \BHII\@. In all cases, the same common trends appear: there is a starburst, which results in an enhancement of the density at the critical radius, causing an increase of the \TDE\ rate around \BHII, followed by a decay on Myr scales. However, the exact moment when the density increases, and its exact peak value, depend on the simulation, showing how small changes (the random seed and therefore the exact location of star formation) in this chaotic system can affect the \TDE\ rate in galaxies.
We note that, since the galaxy hosting \BHI\ is not experiencing strong star formation, the results for \BHI\ are the same in all three re-simulations. Overall, the mean maximal enhancement of the TDE around BH2 in the three simulations is about 140. 

\section{Conclusions}

We assess the \TDE\ rate around \BHs\ using high-resolution hydrodynamical simulations of galaxies during and after a merger with mass ratio 1:4 coupled to the analytical formalism detailed in Section~\ref{tde_rate}. This allows us to track the evolution of the central stellar mass during and after the merger-induced starburst, but also to measure the \TDE\ rate in a realistic, although still idealized environment.

We summarize our findings below:
\begin{itemize}

\item After the first passage below $10 \kpc$, a nuclear starburst promotes an
  enhancement of the stellar density around the \BH\ in the least massive
  galaxy. As a consequence, the \TDE\ rate also increases by up to two orders
  of magnitude for a short duration, and more than one order of magnitude on average.
\item The nuclear starburst produces stellar clumps that scatter the \BH\ and
  modulate the stellar density in its vicinity. The enhancement of the \TDE\
  rate and its duration can therefore vary significantly in different
  realizations of the same process.
\item The environment and \TDE\ rate around the \BH\ in the most massive galaxy
  are rather unaffected by the merger.

\end{itemize}

This confirms that the \TDE\ rate should be larger in galaxies in the final
phases of mergers or the immediate post-merger phase, lasting a few hundreds of
Myr, than in galaxies in isolation. However, large column densities of gas and dust concurrent with the early starbust phases \citep{2017MNRAS.469.4437C,2018MNRAS.478.3056B} can hinder detection of TDEs; whereas the column density decreases in the post-merger phase allowing for easier TDE detection. This picture is independent of the stochastic
behaviour of the star formation process in such a clumpy and turbulent
interstellar medium. However, the exact details of the \TDE\ enhancement, and
the moment it happens, change due the small-scale turbulent dynamics (here mimicked by our perturbed
re-sampling of our stochastic model for star formation), the exact set-up of the initial conditions, and additional parameters, e.g. the existence of a pre-existing cusp, or a different initial gas distribution may modulate the results. 
This is the first
study of \TDE\ rates using hydrodynamical simulations to track how the stellar
profile is modified by star formation and external processes. We stress that this is a proof-of-concept experiment, since we have only explored one particular merger.  Future work will
expand to cosmological simulations.

\section*{Acknowledgements}
MV, YD and HP acknowledge support from the European Research Council (Project no. 614199, `BLACK'). BB is supported by membership from Martin A. and Helen Chooljian at the Institute for Advanced Study. We also thank the anonymous referee for the time taken to carefully read and improve our manuscript. This work was granted access to the HPC resources under the allocations A0020406955 and A0040406955 made by GENCI. This work has made use of the Horizon Cluster hosted by the Institut d'Astrophysique de Paris; we thank Stephane Rouberol for running smoothly this cluster for us.

\label{lastpage}
\end{document}